 \documentstyle[12pt]{article}
\newcommand{\be}{\begin{equation}}
\newcommand{\ee}{\end{equation}}
\newcommand{\bea}{\begin{eqnarray}}
\newcommand{\eea}{\end{eqnarray}}

\newcommand{\th}{\theta}

\newcommand{\half}{{1 \over 2}}

\input epsf

\begin{document}
\begin{center}{\bf \Large      
Exact current-current Green functions in  
strongly correlated 1D systems with impurity.
\footnote{Contribution to the German-Israel winter school in strongly
correlated electron systems, Feb. 21-28, 1997}
}\end{center}
\begin{center} {\it Sergei Skorik} \end{center}
\begin{center} {\it Physics Department, Weizmann Institute for Science, 
Rehovot, Israel}
\end{center}

\vspace{1cm}

{\bf Abstract:} We derive an exact expression for the Kubo conductunce
in the Quantum Hall device with the point-like intra-edge
backscattering. This involves the calculation of current-current
correlator exactly, which we perform using form-factor method.
In brief, the full set of intermediate states is inserted
in the correlator, and for each term the closed mathematical
expression is obtained. It is shown that by making a special choice
of intermediate states in accordance with the hidden symmetries
of the model, one achieves  fast convergence of the series,
thus proving the form-factor approach to be especially powerfull. \newline
This review is based on the joint work with H.Saleur and F.Lesage.

\section{Quantum Hall bar with backscattering and its \\
quantum field theory representation. }

Strongly correlated  fermionic systems revealing 
non-fermi-liquid behavior  is one of the most delicate and
less studied subjects in mesoscopic physics, mainly because
the variety of techniques based on
the perturbation theory are not available in the domain of strong 
interactions.  
One of the commonly used ways to test such systems is to measure the 
conductance as a function of
voltage and temperature. The behavior of the conductance
in the scaling regime of low $V,T$ can to some extent characterize  
non-fermi-liquid state of the system. On the theoretical level, one can start
with some hypothetical effective model, like Luttinger model for 1D
electrons, and show that it correctly accounts for the observed physics. 

It is believed that a pure Luttinger liquid is realized on the edges
of 2D electron gas in the fractional quantum Hall regime. 
In the number of recent theoretical and experimental works the following 
problem has been studied. One takes 2D electron gas in the fractional Qunatum
Hall regime with $\nu=1/3$ in the two-terminal source-drain geometry
shown in Fig. 1 and deforms by static gating potential the narrow
region in the middle, so that the boundaries of electron gas come close
to each other and start to interact.
The quantity of interest is the backscattering 
current $I_{t}(V,T,\omega,V_G)$, or the differential conductance
$\sigma=dI/dV$. The first experimental data for $I_t$ was taken  
by Milliken, Umbach and Webb\footnote{Sol. State Comm. 97 (1996) 309}.
Somewhat more precise experiment was done later by Chang, Pfeiffer and West
\footnote{Phys. Rev. Lett. 77 (1996) 2538} who used cleaved edge
overgrowth technique to measure the tunneling current from the metal
to the Quantum Hall edge.

\begin{figure}
\epsfxsize=100truemm
\centerline{\epsfbox{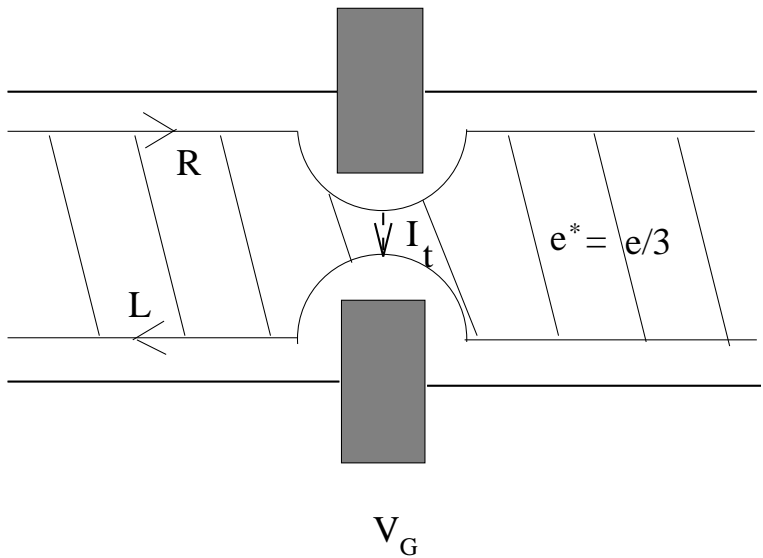}}
\caption{Quantum Hall bar with backscattering between the edge states.}
\end{figure}

To consider the problem
more closely, let us recall first some information about the quantum
Hall bar without gating.
It is known that when one applies a small voltage difference ($V\ll\omega_c$)
to the source
and drain, the current will be carried by the gapless chiral
boundary excitations. Somewhat loose, but nevertheless correct line
of arguments for demonstrating this 
relies on the fact that under Quantum Hall conditons
electrons form incompressible droplet. The change of its volume will
cost a huge amount of energy, while the excitations of arbitrary small
energy can be produced on the boundary of the droplet, by creating some
local lump. It was shown by Wen\footnote{
Phys. Rev. B41 (1990) 12838} that the low-energy edge excitations
can be described by the following effective
(1+1)-dimensional Hamiltonian density:
\be  
{\cal{H}}_0={1\over 8\pi}[\pi^2(x)+(\partial_x\phi)^2]
\label{HAM}
\ee
where $\pi(x)$ is the momentum conjugated to $\phi$.
The field $\phi(x,t)$ represents roughly speaking normal fluctuations of
the edge at point $x$ along the edge and can be factorized
as $\phi(x,t)=\phi_L(t+x)+\phi_R(t-x)$, each component describing one of
the two edges of the bar. The Hamiltonian (\ref{HAM}) represents the energy
density
of such fluctuations and is similar to the Hamiltonian
describing elastic medium. Note that $\phi_L$ and $\phi_R$ are independent
fields, meaning that excitations on two edges are not interacting with
 each other.
The point is that not all eigenstates of the Hamiltonian (\ref{HAM}) 
are real physical excitations observed in the Quantum Hall system. The Hilbert
space of eigenstates of (\ref{HAM}) is  sort of a ``playground'' which 
consists of the vacuum $|0\rangle$
and many-particle bosonic states $\phi_L(x_1)...\phi_R(x_N)|0\rangle$
out of which one must select the subspace of {\it physical} states. The one-electron
excitation is the following state:
\be
\Psi^+_L|0\rangle = :e^{i\phi_L/\sqrt{\nu}}:|0\rangle, \qquad
\Psi^+_R|0\rangle = :e^{i\phi_R/\sqrt{\nu}}:|0\rangle
\label{ELECTR}
\ee
and the electric charge along the edge is given by the following
operators:
\be
Q_L=e\int :\Psi_L^+\Psi_L:=e\sqrt{\nu}\int \partial_x\phi_L,
\qquad
Q_R=e\int :\Psi_R^+\Psi_R:=-e\sqrt{\nu}\int \partial_x\phi_R
\label{DENS}
\ee
It is said that $\Psi^+_L (\Psi^+_R)$ creates an electron on the left
(right) edge, which is represented by the commutation
relation $[\Psi_L, Q_L]=e\Psi_L$. 
The correct anticommutator $\{\Psi_{L,R}^+(x), \Psi_{L,R}(x')\}=\delta(x-x')$
is ensured by the normal ordering of exponential and by $[\phi_{L,R}(x), 
\phi_{L,R}(x')]=\pm i\pi sign (x-x')$ 
(such a representation of fermions in terms of free bosonic
states is known as {\it bosonization}).

 Beside electrons, there exist
quasiparticle excitations with the fractional charge $e^\ast=\nu e$.
 The quasiparticles are created by the operators
similar to those of electrons, with $\nu\to 1/\nu$ in the exponent:\footnote{
Note that we tacitly assume everywhere $\nu=1/3$, however, the above
results are valid for $\nu^{-1}$ odd integer.}
\be
\Psi^+_{qL}|0\rangle = :e^{i\sqrt{\nu}\phi_L}:|0\rangle, \qquad
\Psi^+_{qR}|0\rangle = :e^{i\sqrt{\nu}\phi_R}:|0\rangle
\label{QUASI}
\ee

The gating of the bar allows for the scattering of electrons and quasiparticles
from one edge to another through  intermediate impurities in the region
where two edges are close to each other. Under the assumption that
the scattering takes place at one point, $x=0$, we write down the most
general term decsribing many-particle backscattering processes:
\be
H_{bs}=\sum_{n=1}^\infty b_n[\Psi_{qL}^+\Psi_{qR}]^n(0) + h.c.
=\sum_{n=1}^\infty b_n\cos[n\sqrt{\nu}(\phi_L-\phi_R)]_{x=0}
\label{BS}
\ee
In particular, the $n=1$ term is just one quasiparticle tunneling
process, and $n=1/\nu$ term is electron tunneling. The form of (\ref{BS})
suggests that
quasiparticles are more fundamental objects than electrons, the electron
tunneling simply being the coherent tunneling of $1/\nu$ quasiparticles.
Further, the renormalization group analysis shows that in the limit
of small energy
scales (large distances) the correlation functions of the $H_0+H_{bs}$
system are exactly the same as the low energy correlation functions 
of the Hamiltonian
\be
H={1\over 8\pi}\int_{-\infty}^{+\infty}dx
[\pi^2(x)+(\partial_x\phi)^2]
+b_1\cos[\sqrt{\nu}(\phi_L-\phi_R)]_{x=0}
\label{FULLHAM}
\ee
This is the basic theoretical model, derived in the pioneering work
by Wen\footnote{Phys. Rev. B44 (1991) 5708}. 
 We show in the following sections how to obtain
{\it exactly}  expression for the linear response $\sigma(\omega,b_1)$
for $T=0$ using Kubo's formula and the model (\ref{FULLHAM}).

\section{Mapping to a half-line.}

The Hamiltonian (\ref{FULLHAM}) is a particular form of  1D field theory
with impurity. After making a canonical transformation on the fields
one can rewrite (\ref{FULLHAM}) as a theory on a half-line with a boundary
interaction. The reason for doing this will become clear later. Introduce
\bea
\phi_1(x+t)&=&{1\over\sqrt{2}}\left[\phi_L(x,t)+
\phi_R(-x,t)\right]\nonumber \\
\phi_2(x+t)&=&{1\over\sqrt{2}}
\left[\phi_L(x,t)-\phi_R(-x,t)\right] \label{last:remii}
\eea
As a result, field $\phi_1$ decouples from $\phi_2$ as a free field,
while $\phi_2$ bears all the interactions. \newline
\underline{\it Remark:}
With the backscattering
of the form   (\ref{BS}) the total electric charge on two edges,
$Q_L+Q_R$, is still conserved, although $Q_L$ and $Q_R$ are not conserved
separately. It is easy to check that  $Q_L+Q_R$ is expressed purely in terms
of field $\phi_1$, while $Q_L-Q_R$ depends only on $\phi_2$.  

Further, one folds the line  by defining a new field, $\Phi=
\Phi_L+\Phi_R$
on a half-line as follows:
\bea
\Phi_L(x,t)&=&{1\over 2\sqrt{\pi}}\phi_2(x+t), \quad x>0, \nonumber \\
\Phi_R(x,t)&=&{1\over 2\sqrt{\pi}}\phi_2(-x+t), \quad x>0.\label{last:remiii}
\eea
So, we arrive to the boundary sine-Gordon Hamiltonian
\be
H={1\over 2}\int^{\infty}_0 dx
[\Pi^2(x)+(\partial_x\Phi)^2]
+b \cos{\sqrt{8\pi\nu}\over 2}\Phi(0),
\label{BSGHAM}
\ee

\section{Definition of the boundary state.}

\begin{figure}
\epsfxsize=75truemm
\epsfysize=75truemm
\centerline{\epsfbox{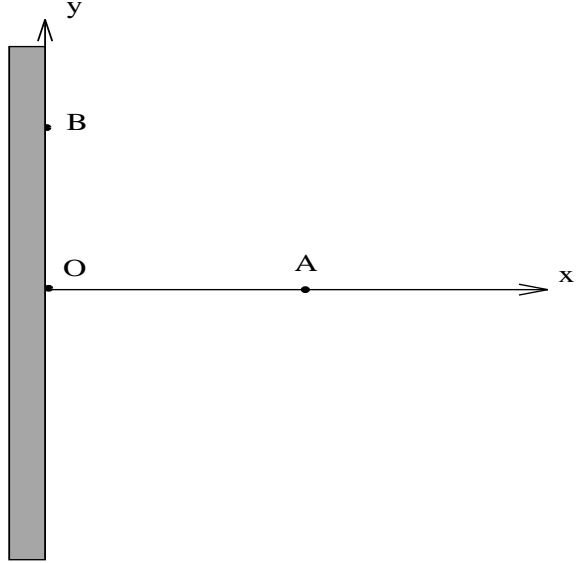}}
\caption{The geometry of space-time.}
\end{figure}

We will work on the half-plane which geometry is shown in Figure 2.
Based on the euclidean duality,
there are two alternative ways to introduce the Hamiltonian picture.
First, one can take $x$ to be euclidean time. In this case the equal time 
section is an infinite line $x=$const, $y\in(-\infty,\infty)$. Hence the
associated space of states is the same as in the bulk theory.
In our case the bulk theory
related to the Hamiltonian (\ref{BSGHAM}) is just a massless free
boson on the line:
\be
H={1\over 2}\int^{\infty}_{-\infty} dx
[\Pi^2(x)+(\partial_x\Phi)^2],
\label{FREEHAM}
\ee
obtained from (\ref{BSGHAM}) by changing the roles of space and time.
The boundary term at $x=0$ dissapears from the Hamiltonian, but appears as 
 the ``time boundary,'' or initial condition at $x=0$
which is described by the {\it boundary state} $|B\rangle$ (a particular
state from the bulk Hilbert space). The correlators are expressed as
\be
\langle O_1(x_1,y_1)...O_N(x_N,y_N)\rangle = 
{\langle B|{\cal T}_x[O_1(x_1,y_1)...O_N(x_N,y_N)]|0\rangle
\over \langle 0|B\rangle}, \label{last:viewI}
\ee
where $O_i(x,y)$ are the Heisenberg local field operators
\be
O_i(x,y)=e^{-xH}O_i(0,y)e^{xH},
\ee
and ${\cal T}_x$ means x-ordering. 

Alternatively, one can take the direction along the boundary to be the time.
In this case boundary appears as a boundary in space, and the Hilbert space
of states is associated with the semi-infinite line $y=$const, 
$x\in[0,\infty)$.
The correlation functions of any local fields $O_i(x,y)$ can be
computed in this picture as the matrix elements
\be
\langle O_1(x_1,y_1)...O_N(x_N,y_N)\rangle = 
{\langle 0||{\cal T}_y[O_1(x_1,y_1)...O_N(x_N,y_N)]||0\rangle
\over \langle 0||0\rangle}, \label{last:viewII}
\ee
where $||0\rangle$ is the ground state of the boundary Hamiltonian,
$O_i(x,y)$ are understood as the corresponding Heisenberg operators
\be
O_i(x,y)=e^{-yH_B}O_i(x,0)e^{yH_B},
\ee
and ${\cal T}_y$ means y-ordering.

The equality of expressions (\ref{last:viewI}) and (\ref{last:viewII})
can be understood as a definition of the boundary state, which is chosen
such as to provide the equivalence of correlators. Ghoshal and Zamolodchikov
\footnote{
S.Ghoshal, A.Zamolodchikov, Int. J. Mod. Phys. A9 (1994) 3841} found
the generic form of the boundary state 
for the special class
of models, called {\it integrable}, including the model of our interest
(\ref{BSGHAM}):
\bea
|B\rangle=|0\rangle+\sum_{N=1}^{\infty}&&\int_{-\infty<\th_1<\dots<\th_N<\infty}
R^{a_1}_{\bar{b}_1}\left({i\pi\over 2}+\th_1\right)
 \dots
R^{a_N}_{\bar{b}_N}\left({i\pi\over 2}+\th_N\right) 
 \nonumber \\
&&A_L^{a_N}(\th_N)\dots A_L^{a_1}
(\th_1)A_R^{b_1}(\th_1)\dots A_R^{b_N}(\th_N)|0\rangle
\label{last:bsI}
\eea  
This expression must be understood as follows: indexes $a_i,b_i$ label
the full set of particles which scatter off the boundary with the
{\it boundary reflectiom matrices} $R^a_b(\theta)$, and $\bar{b}$ is 
an antiparticle corresponding to a particle $b$. 
The summation over the particle indices $a,b$ is assumed. The rapidity
$\theta$ parametrizes the momentum of particles as
\be
E_L=-P_L=\mu e^\theta, \qquad E_R=P_R=\mu e^\theta,
\label{eq:enmom}
\ee
with $\mu$  an arbitrary
mass scale,
and $A_{L,R}^{a,b}(\theta)$ are the operators that create 
from the vacuum left and right moving particles $a,b$ at rapidity $\th$.
From the structure of the state (\ref{BS}) one can infer what are some of the 
distinguished properties of the integrable theories:  particles scatter
off the boundary one by one (without particle production) 
with their energy preserved.

\section{Outline of the method.}

We are interested in the   AC conductivity at zero temperature 
which is given by Kubo's formula
\be
{
\sigma(\omega)=\int_0^\infty dx\int_0^\infty
\langle 0|j(0)j(t+ix)|0\rangle e^{-i\omega t}dt
}\label{last:kubo}
\ee
The net current $j$ here is difference of the currents in left and right
edges, $j=\partial_t\phi_L-\partial_t\phi_R$. Thus, one has to evaluate
the correlator $\langle 0|\partial_t\phi_L(0)\partial_t\phi_R(x)|0\rangle$
for the model (\ref{FULLHAM}) or, equivalently, the correlator
$\langle B|\partial_t\Phi(0)\partial_t\Phi(x)|0\rangle$ for the
model (\ref{BSGHAM}). In the form-factor approach that we adopt   
one  inserts the full set of intermediate n-particle
states in the correlator $\langle B|j(0) j(x)|0\rangle=\sum_n
\langle B|j(0) |n\rangle\langle n|j(x) |0\rangle$. Then the problem is to
find the form-factors $\langle n|j(x)|0\rangle$ and perform the summation
over the complete set of intermediate states. For obvious reasons, it is
preferred for the intermediate states
to take the eigenstates of the Hamiltonian (\ref{BSGHAM}).

Let us comment on this last point. In the massless free field theories 
there is a freedom to choose a complete basis of particle eigen-states.
Often,  for the basis one takes  plane waves, as in the canonical
quantization procedure. However, other representations of particle
states can be obtained if one regards the massless  theory as a limit
of certain interacting massive theory. Then, upon switching off 
{\it bulk} interactions
and taking the massless limit, one obtains certain massless particle
states, remniscent of  massive particles, depending from which
massive field theory one approaches the massless limit. Of course, one
can constract these states as wave packets of certain energy 
from the plane waves due to the degeneracy of the Hilbert space, but
for this one has to know the matrix elements between plane waves and such
states, which is not easy to obtain.

Having this in mind, we shall consider the theory (\ref{BSGHAM}) as a massless
limit of the  massive sine-Gordon model:
\be
H_{SG}={1\over 2}\int^{\infty}_{-\infty} dx
[{\tilde{\Pi}}^2(x)+(\partial_x\tilde{\Phi})^2 + 
m\cos\sqrt{8\pi\nu}\tilde{\Phi}]
+\tilde{b} \cos{\sqrt{8\pi\nu}\over 2}\tilde{\Phi}(0)
\label{BULK}
\ee
It was shown by Ghoshal and Zamolodchikov in the work cited above
that the model (\ref{BULK}) possesses infinite set of 
mutually commuting local conserved charges. The eigenstates of (\ref{BULK})
are well-known massive particles, called {\it kinks}, {\it antikinks}
and their bound states -- {\it breathers}, which diagonilize
 the conserved
charges. Correspondingly, in the limit $m\to 0$ the conserved charges
of (\ref{BULK}) become the conserved charges for the model (\ref{BSGHAM})
expressed in terms of the massless field $\Phi$,  their eigenstates
 being massless limit of kinks, antikinks and breathers. The major
advantage of working in the basis of massless kinks and antikinks is that
their boundary scattering matrices are just 2x2 matrices, obtained 
by Ghoshal and Zamolodchikov exactly. Thus, the boundary state $|B\rangle$
is significantly simplified and given explicitly by  expression 
(\ref{BS}). The integrable models techniques allows also to find exactly
the form-factors for the multiparticle kink states.
As it is shown below, the summation
over the  states presents no difficulties, for the series converge
rapidly. 

The use of relation between (\ref{BSGHAM}) and its massive analogue
(\ref{BULK}) for determining  proper particle states,
and the hidden symmetries of  (\ref{BSGHAM}) is the
key feature of our approach. Our results are non-perturbative since we
do not assume the smallness of the coupling constants $b$ or $\nu$,
the expansion parameter being rather a volume of the multi-particle
phase space. The hidden symmetries provide  the fast convergence
of the multi-particle exansion by suppressing the multi-particle creation
processes.

\section{Calculation of correlation functions}

In what follows we will stay with the first representation of correlators,
through the boundary state. This means that in Figure 2 coordinate 
$x$ is imaginary time and $y$ is
space coordinate, $z=x+iy$. Time translation is performed by
the operator $T=\exp(xH)$, with $H$ being the bulk Hamiltonian (\ref{FREEHAM}).

We need compute the following matrix element:
\bea
 \langle B|\partial_x\Phi(x_1,y_1)\partial_x\Phi(x_2,y_2)|0\rangle
&=&
\langle B|\partial_z\Phi_R(z_1)\partial_{\bar z}\Phi_L(\bar{z_2})|0\rangle
\nonumber \\
&+&
\langle B|\partial_{\bar z}\Phi_L(\bar{z_1})\partial_z\Phi_R(z_2)|0\rangle
\label{cor}
\eea
where $\Phi$ is a massless field $\Phi=\Phi_L(\bar{z})+\Phi_R(z)$
and $|B\rangle$ denotes the  boundary state 
of the sine-Gordon model (\ref{BSGHAM}). 
Because $|B\rangle$ has chirality zero,  products of the fields
of the same chirality do not contribute to the right hand side of eq.
(\ref{cor}).

Substituting the boundary state (\ref{BS}) into (\ref{cor})
and using the fact that left (right) moving field acts only
on left (right) moving particles, we obtain the following expansion
in terms of the form-factors:
\bea
&&\langle B|\partial_x\Phi(x_1,y_1)\partial_x\Phi(x_2,y_2)|0\rangle=
\langle 0|\partial_x\Phi(x_1,y_1)\partial_x\Phi(x_2,y_2)|0\rangle+
\nonumber \\
 &+&
\sum_{N=1}^\infty \sum_{a_i,b_i}
\mu^2\int_{-\infty<\theta_1<...<\theta_N<\infty}
\left[\prod_{i=1}^N{d\theta_i\over 2\pi}R^{a_i}_{\bar{b}_i}
({i\pi\over 2}+\theta_i)\right]
\langle 0|\Phi_L(0)|A^{a_1}(\theta_1),...,A^{a_N}(\theta_N)\rangle_L 
\nonumber \\
&\times&\langle 0|\Phi_R(0)|A^{b_1}(\theta_1),...,A^{b_N}(\theta_N)\rangle_R
\left(\sum_{i=1}^N e^{\theta_i}\right)^2
\left(e^{-(x_1+x_2+iy_1-iy_2)\mu \left(\sum_{i=1}^Ne^\theta_i\right)} + 
{\rm c.c.}\right) \nonumber \\
&\equiv& G_0(z_1-z_2)-\sum_{n=1}^\infty I_{n} \label{corI}
\eea

The first term in (\ref{corI}), $G_0$, is the Green's function of the 
free massless scalar (\ref{FREEHAM}) on the plane:
\be
G_0(z,\bar{z})={1\over 4\pi}\left({1\over z^2}+{1\over \bar{z}^2}\right) 
\ee
Obviously, it depends only on the difference of the points, while
the rest of the terms $I_n$ depend on $x_1+x_2$ as a consequence of
the translational invariance breakdown on the half-plane.

The particle spectrum in (\ref{corI}) is determined by the model
(\ref{BULK}) for $\nu=1/3$ and consists of a massless kink $A^+(\theta)$,
anti-kink $A^-(\th)=\bar{A}^+(\th)$, and a massless breather $A^0(\th)$.
Corresponding  boundary reflection matrices are:
\bea
R^0_0(\th)&=& -\tanh{1\over 2}(\th-\th_B-{i\pi\over 2})     \nonumber \\
R^{\pm}_{\mp}(\th)&=&e^{\sqrt{8\pi/3}}D(\th-\th_B), \nonumber \\
R^{\pm}_{\pm}(\th)&=&e^{-\sqrt{8\pi/3}}D(\th-\th_B), \label{eq:Rpm}
\eea
\be
D(\vartheta)={1\over 2\cosh(\vartheta-{i\pi\over 4})} 
{\Gamma({3\over 8}-{i\vartheta\over 2\pi})\Gamma({5\over 8}+
{i\vartheta\over 2\pi})
\over\Gamma({5\over 8}-{i\vartheta\over 2\pi})\Gamma({3\over 8}+
{i\vartheta\over 2\pi})}
\label{last:R}
\ee
The impurity coupling constant $b$ enters the correlator only through
the above scattering matrices, and we parametrized $b(T_B)$ in terms
of the boundary temperature $T_B=\mu e^{\theta_B}$. The $b=0$ point
corresponds to $\th_B=-\infty$, while $b=\infty$ corresponds to
$\th_B=\infty$. The momentum and boundary temperature enter
the scattering matrices in the dimensionless combination $P/T_B$,
which means that $R^a_b$ depend on the difference $\th-\th_B$.
The form-factors $
\langle 0|\Phi_{L,R}(0)|A^{a_1}(\theta_1),...,A^{a_N}(\theta_N)\rangle_{L,R}$
have been obtained by Smirnov\footnote{``Form-factors in completely
integrable models of quantum field theory'', World Scientific 1992}.

The first term,  $I_1$, in (\ref{corI}) is the one-breather contribution.
The corresponding form-factors 
$\langle 0|\Phi_L(0)|A^0(\theta)\rangle_L $ and
$\langle 0|\Phi_R(0)|A^0(\theta)\rangle_R$ are just constants. Explicitly,
we have
\be
I_1={\mu^2\over 2N_0}
\int_{-\infty}^{+\infty}{d\theta\over 2\pi}e^{2\theta}
\left(e^{-(x_1+x_2-iy_1+iy_2)\mu e^\theta} + {\rm c.c.}\right)
\tanh{\theta-\theta_B\over 2} \label{contribI}
\ee
where $N_0$ is some normalization constant of the order of 1.
Changing variables back to momentum, $p=\mu\exp(\theta)$, 
it can be rewritten as
\be
I_1={1\over 4\pi N_0}\int_0^\infty pdp {p-T_B\over p+T_B}
\left(e^{-(z_2+\bar{z}_1)p} + c.c\right)
\label{last:prepr}
\ee
(the arbitrary mass scale $\mu$ dissapeared from the final answer).
Plot of the one-breather contribution to the two-point
correlation function for the points OA ($x_1=y_1=y_2=0$) is 
shown in Figure 3.
The three-breather contribution $I_3$ related to  the form-factors
$\langle 0|\Phi_{L,R}(0)|A^0(\theta_1),...,A^0(\theta_3)\rangle_{L,R}$
is given by
\bea
I_3={\mu^2H_3^2\over 3! N_0}&&\int_{-\infty}^\infty
\prod_{i=1}^3{d\theta_i\over 2\pi}\tanh{\theta_i-
\theta_B\over 2}
\label{contribIII} \\
&\times&\prod_{i<j}^3{|F_{min}(\theta_i-\theta_j)|^2 \over
2(1+\cosh(\theta_i-\theta_j))}\left(\sum_{i=1}^3e^{\theta_i}\right)^2
\left(e^{-(z_2+\bar{z}_1)\mu\left(\sum_{i=1}^3e^{\theta_i}\right)} + 
c.c\right) \nonumber
\eea
where $H_3$ is some constant and $F_{min}$ is some complicated known function
which we do not specify here. The important fact is that the value
of $I_3$ is 100 times smaller than $I_1$. The term $I_2$ is a two-particle
contribution of kink and anti-kink, which is related to the form-factors
 $\langle 0|\Phi_{L,R}(0)|A^+(\theta_1)A^-(\theta_2)\rangle_{L,R}$
(in general, kinks and anti-kinks appear only in pair in the form-factor
expansion). The magnitude of this term
is approximately 20\% of the value of $I_1$. Contributing to $I_3$ is
also the three-particle intermediate state of kink, anti-kink and breather.
It is clear how to obtain the rest of the terms.
We do not list corresponding expressions here, because $I_n$ decrease
very fast with $n$ which makes possible to truncate the series for most
of the purposes.\footnote{Interested reader can see the paper in
Nucl.Phys.B474 (1996) 602}
  Each integral
$I_n$ converges for any finite value of $(x_1+x_2)$, but is divergent
for $x_1=x_2=0$. It is possible to continue analytically our integral
representations of $I_n$ to the boundary domain  $x_1=x_2=0$ \footnote
{S.Skorik, PhD thesis, hep-th/9604174   }.
 
\begin{figure}
\epsfxsize=100truemm
\centerline{\epsfbox{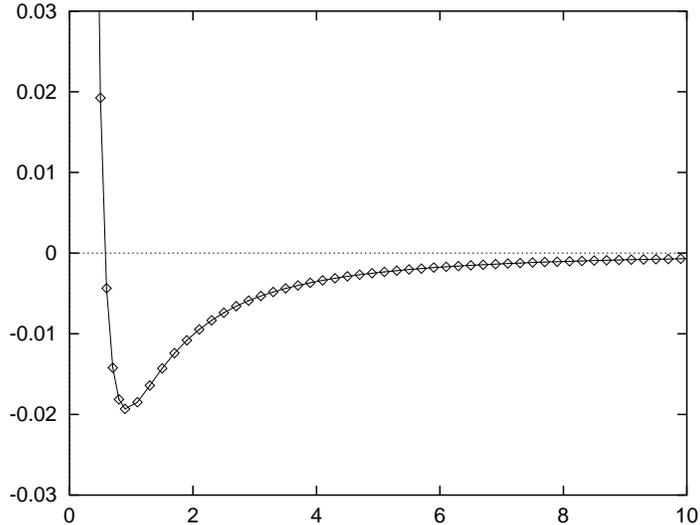}}
\caption{Plot of the one-particle contribution to correlator between
               two points $(0,0)$ and $(x,0)$ as a function of $x$.}
\end{figure}

\section{Scaling from large to small energies.}

The non-perturbative nature of the form-factor approach allows one
to study the behavior of correlation functions with the change of the scale. 
Let us study the behavior of integrals $I_n$ under the dilatation
 $z\to e^\lambda z$. Such a rescaling can be compensated
by the change $\theta_B\to\theta_B+\lambda$ and by the overall
normalization factor $Z(\lambda)=e^{2\lambda}$ to have the integrals
(and hence the correlator) unchanged. Repeating this RG transformation,
we will flow to the UV or IR fixed points $\theta_B=\pm\infty$
(depending whether $\lambda>0$ or $\lambda<0$).
For
such values of $\theta_B$ the hyperbolic tangent factors in the
integrands are equal to $\pm 1$, and the integrals are proportional
 to $\pm 1/z^2$.
On the plot in Figure 3  one can see two regimes:
$\mu z\ll 1$ and $\mu z \gg 1$ when the functions behave as $\pm 1/z^2$
(in other words, far away from the boundary an observer will experience
the fixed IR boundary condition $\Phi(0)=0$, while very close to the
boundary -- free UV boundary condition $\partial_x\Phi(0)=0$).
The non-trivial behaviour at the intermediate scales is due to the presence
of boundary, which introduces a scale $\mu e^{\theta_B}$
corresponding approximately to the position of the deep. Shifting
$\theta_B$ corresponds to the motion of the deep to the right or
left on Figure 3, untill it will go away completely and one of the
regimes will dominate over all scales.

\section{Conductivity.}

The leading contribution to the conductance computed along the lines
of the discussion above  for $\nu=1/3$ is given by
\be
 G(\omega)^{(1)}={1\over 6}-\kappa^2{\pi d^2\over 8}{\rm Re}\tanh\left[
\half\log\left({\omega\over\sqrt{2}T_B}\right)-{i\pi\over 4}\right]
\label{last:deltaGI}
\ee
where $\kappa\approx 3.14 $ and $d\approx 0.1414$.
We plot the function $G(\omega)^{(1)}$ in Figure 4 and emphasize that it
 reproduces the shape of the exact condactance to a very high accuracy
{\it at any value of coupling }$T_B(b)$. The higher order terms 
add some slight corrections to the shape of $G(\omega)^{(1)}$ which are
comparable with the available accuracy of experimental measurements
(obviously, when probing the conductance in experiment one expects the 
agreement
with the theory for sufficiently small $\omega$; 
for example when $\omega$ approaches
the magnetic gap $\omega_c$ the validity of predictions
based on the Wen's theory (\ref{FULLHAM}) breaks down).

\begin{figure}
\epsfxsize=75truemm
\centerline{\epsfbox{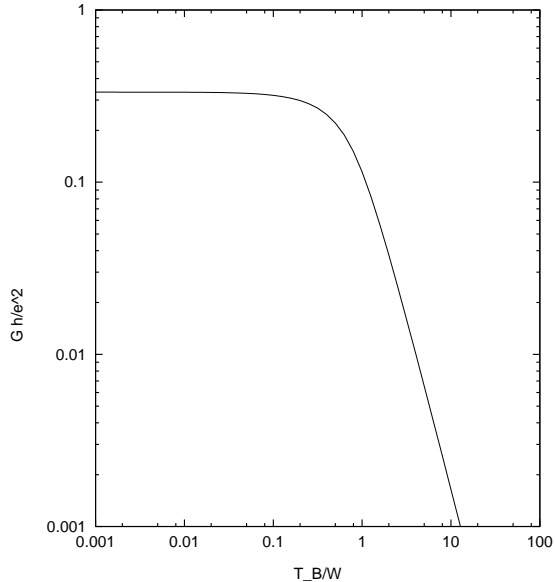}}
\caption{Frequency dependent conductance at $T=0$.}
\end{figure}

\section{Discussion.}

In this methodological review we described a quantum-field theoretical
approach to certain impurity problems. The natural question to ask
is how generic is this approach, to what extent its applicability is 
restricted.
 The above method works for the integrable models,
the integrability being a strong constraint. For example, adding another
impurity (or another gating potential nearby in the set-up described here)
will destroy integrability of the model. However, in such cases
one can develop a perturbative expansion over the integrable model
by using the ideas given above, in particular by making use of the
quasiparticle basis. We hope that such expansions will give better
results than perturbations over the free field theory with the plain wave
basis. The models that have been so far treated by the above technique
or can be treated in principle include Kondo model and its multi-channel
analogue, spin-boson model of dissipative quantum mechanics, quantum Hall
bar with one or two constrictions, quantum dot and physics of 
coulomb blockade.   

\end{document}